# Experimental investigation of Lord Kelvin's isentropic thermoelastic cooling and heating expression for tensile bars in two engineering alloys

S. J. Burns [a,b], Christine E. Pratt[b], Joseph Carrock[c], Jean-Philippe Gagnon[d] and A. B. Sefkow[b,e,f,g]

[a]Materials Science Program, College of Arts, Science and Engineering, University of Rochester, Rochester, NY, USA; [b]Department of Mechanical Engineering, University of Rochester, Rochester, NY, USA; [c]Field Application Engineering, Telops, Geneva, NY, USA; [d]Systems Engineering, Telops, Quebec, QC, Canada; [e]Laboratory for Laser Energetics, University of Rochester, Rochester, NY, USA; [f]Department of Physics and Astronomy, University of Rochester, Rochester, NY, USA; [g]Department of Computer Science, University of Rochester, Rochester, NY, USA

**ABSTRACT**

Solids change temperature when rapidly and elastically stressed. The effect proposed by W. Thomson, who was ennobled as Lord Kelvin, is adiabatic thermoelastic cooling or heating depending on the sign of the stress. A highly sensitive radiometrically calibrated, cooled infrared camera was employed to measure temperatures both decreasing and increasing. Temperature measurements made from the reversible, elastic part of the stress–strain curve during rapid uniaxial tensile loading and unloading were investigated. The quasi-isentropic cooling temperature from the stress loading curve is recovered by heating after the specimen fractures when the load is released. These measurements establish for the first-time isentropic temperature recovery. The materials tested are an AISI 4340 steel and an aluminium 2024 alloy. Measurements of the stress cooling are $-0.65 \pm 0.05$ K/GPa for steel and $-1.8 \pm 0.3$ K/GPa for aluminium alloy. The thermoelastic heating is $-1.2 \pm 0.3$ K/GPa for steel and $-1.8 \pm 0.2$ K/GPa for aluminium. These values are compared to Thomson's theoretical thermoelastic prediction; comparisons are also made to equations of state tables. The experimental values are below theoretical predictions. The quasi-isentropic, elastic part of the temperature change is fully recovered after extensive plastic deformation by the fracture stress release wave.

**CONTACT** S. J. Burns stephen.j.burns@rochester.edu Materials Science Program, College of Arts, Science and Engineering, University of Rochester, Rochester, NY 14627, USA; Department of Mechanical Engineering, University of Rochester, 236 Hopeman Hall, Rochester, NY 14627, USA

## 1. Introduction

An adiabatically compressed gas heats; a rapidly expanded gas cools. A solid tensile bar quickly and elastically stressed also changes temperature. The quantitative measurements of temperature changes due to elastic, reversible stresses in uniaxially stressed, solid, tensile bars are described below.

W. Thomson's contributions to the basic laws of thermodynamics are well-known [1, 2]. He investigated the role of energy balances, including the heat terms, on thermoelastic and thermomagnetic properties of materials. The temperature changes in stressed solid materials follows from references [1, 2] which are part of his foundational works. The stress driven isentropic temperature expressions were verified in solids at that time by Joule [3]. The thermal elastic effects although small, fast, and difficult to measure, were measured by the exceptional experimental investigations of Joule who not only studied metals but also pine wood, Indian and vulcanised rubbers, etc. Thermoelastic behaviour has been subsequently studied with experimental measurements [4–6] to establish the soundness of Thompson's thermoelastic predictions especially in metals, rubbers and polymer based laminated composites [7–9].

The close relationship of irreversible work and temperature in mechanical testing at high strain-rates is a very active area of research especially using Taylor-Quinney ratios [10] to describe thermal effects in stress vs strain curves. The adiabatic ratio of heat to mechanical cold work, changes the material's temperature as noted in recent descriptions [11–19] of a material's thermal behaviour. Shear driven, near adiabatic behaviour which includes micro-structural changes [20] were initially observed using a calorimeter; the analysis originally concentrated on torsional test bars with only shear plastic deformation. The Taylor-Quinney ratio of the heat, proportioned to the cold work in the plastic part of the stress vs strain curve, measures and analyzes temperature changes [10, 19, 21, 22]. Tensile test elastic temperature changes come prior to the material's temperature change by irreversible yield-deformation and are thus necessarily included in Taylor-Quinney ratios.

Solid materials are far more complex than gases or liquids since solids support shear stresses. Solids have six independent stresses [23], yet heat capacity is often described simply by either constant pressure or constant volume. Pressure or volume considerations versus stress or strain variables in uniaxial tensile bars are different concepts since uniaxial stresses in general contain shear stresses which are nonexistent in systems that are only pressurised. Shear is important in tensile testing [24–26] as it governs the material's properties from irreversible dislocation motion and/or twinning deformation behaviour. Shear stress is a major consideration in tensile testing as it is the basis for dislocation plasticity but is it important in the thermoelastic behaviour? Consider the following: elastic moduli are typically considered as either isothermal or adiabatic elastic: the isentropic restriction implies that the temperature will change when rapidly stressing the material. Thermoelastic stress coupling is dependent on the normal stress components in the elastic modulus tensor; the shear moduli between isothermal and isentropic values shows no differences at zero shear stress. Thus, the shear stress should make no contribution to the thermoelastic effect. However, there is heat involved in the isothermal behaviour of sheared solids: it has been shown [27] that the ratio of heat to elastic shear energy in isothermal elastic longitudinal compression is a number that depends only on the temperature of the shear compliance. Thus, shear stresses are coupled into the thermoelastic behaviour of sheared material; it is unknown how these stresses contribute to elastic temperature changes.

Tensile tests [11–13] are one of the most widely used experimental measurements to determine mechanical properties of materials. These tests are often made over a wide selection of temperatures while the samples are considered near adiabatic at strain-rates of $10^{-1}$/s or higher; the strain rate in the release stress wave from specimens after fracturing exceeds $10^{+3}$/s, again dependent on the test sample size. Temperature changes from fracture release stress waves are rarely shown and are not discussed in any studies of thermoelastic deformation that we are knowledgeable about. We know of no release wave elastic measurements in the literature with temperature changes due to elastic deformation as might be found using an IR camera, although, the release waves are sometimes seen in experimental thermal measurements.

Aluminium and ferrous materials have pressure-based equations of state, EOS [28] that theoretically predict temperatures to extreme pressures on adiabatic lines: Sesame Table #3720 is for aluminium up to $10^{18}$ Pa and Sesame Table #2140 is for iron up to pressures of $10^{16}$ Pa. Both tables are for temperatures of 1000's of degrees K and higher. Sesame's EOS are thermodynamic surfaces of entropy, pressures, specific volumes, and temperatures for a wide selection of materials. The data measured in the experiments that follow are for temperature changes due to uniaxial stresses which are compared to the isentropic surfaces on the thermodynamic EOS from Sesame and LEOS tables. The temperature changes predicted by the two Sesame and one LEOS tables are at the very low-pressure part of the tables and are in tension.

Isothermal and isentropic behaviour are limits that are difficult to achieve experimentally. The term quasi-isentropic behaviour is used in our experimental programme. The material is fully elastic, but the speed of the testing may allow for some limited heat flow. Heat flow is discussed and quantified in our Section 5, *Analysis for Thermo-Elastic Stress Cooling and Heating*. Analysis of experimental quasi-isentropic behaviour from cooling and heating follows in Sections 2 and 3.

## 2. Mechanical and thermal experimental measurements

### 2.1. Mechanical measurements

The Materials Test System, Criterion Model 43, load frame running Test Works 4 software was used for mechanical data acquisition at 100 Hz. The tests were conducted with an applied cross-head speed of 0.423 mm/s. The engineering stress versus engineering strain graphs are recorded with the aid of the system's extensometer. The tensile bars were purchased from Laboratory Devices Co. (Loomis CA, USA) and were made for thermo-elastic tests. The materials tested are 2024 T351 aluminium bars and bars of AISI 4340 steel. The 2000 series aluminium is a wrought, heat-treatable aluminium-copper-magnesium alloy with Guinier-Preston precipitates that harden the alloy. It was used in the tempered condition with high yield strength and ultimate stress and excellent ductility. The steel's test bar is a low carbon ferrous alloy of nickel-chrome-molybdenum in a normalised state. This alloy is extensively used for deep hardening, martensitic, high strength steels. The tensile bars have a gage length of 80 mm. We used an extensometer of about 50 mm gage length for measuring engineering strain. Figure 1 is a photograph of the load frame, a steel tensile specimen, the IR camera plus the load cell. Figure 1(a) shows the camera's overall alignment for recording the temperature in the test sample and its position relative to the MTS instrument's load frame. Two bars of each alloy were coated with LabIR® thermographic black spray paint (TIMI Creation, Plzeň, Czech Republic) to increase thermal emissivity and reduce reflectivity as seen in Figure 1(b). Application of high emissivity paint improves the collectability of radiometrically accurate temperature results without post-processing of the data.

Elastic stress rates and stress versus time curves are described below and in Figures 2 and 3. Our results were recorded versus time because the instruments used were asynchronous. We have also post-processed about 10% of our experimental measurements to generate synchronous measurements of temperature and stress. Figure 2(a) is a plot of the mechanical stress versus time curve for the aluminium alloy. Time and strain are closely related to the MTS's constant cross-head rate with an in-line machine compliance. The elastic response seen on the left at the start of the graph establishes where the material is reversible. The specimen in the low stress region remains linear elastic for about 3 s. Also, note that it takes about a second for the cross-head to fully engage the specimen. We measured Young's modulus using the extensometer in the elastic region to be 72.4 GPa for this alloy. Literature data [29] from the manufacture suggests a modulus of 73.1 GPa. Figure 2(b) shows the increase in elastic stress versus time for this sample. It establishes the elastic loading rate for the aluminium bars as 157 MPa/s. The loading rate taken from Figure 2(b) is well below the yield stress for the alloy but with times long enough to establish full mechanical contact in the load-train fixtures.

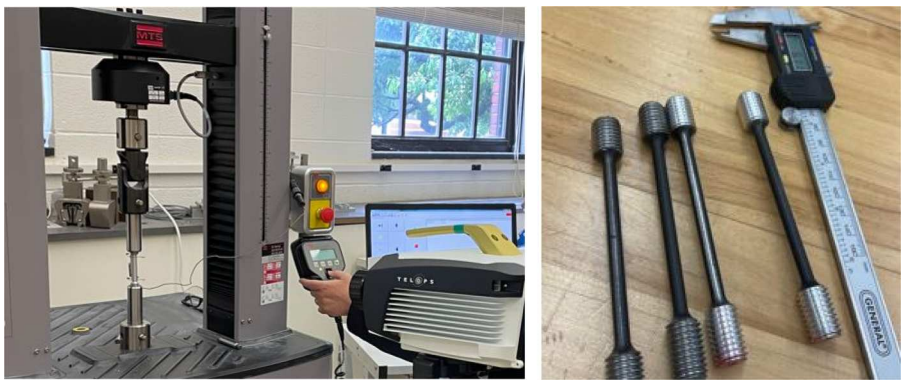

**Figure 1.** (a) A photograph of the test frame, IR Camera, and an aluminium test bar with a clip gage extensometer. (b) The image of aluminium on the right and steel coated test bars. The test bars have been IR coated to increase emissivity.

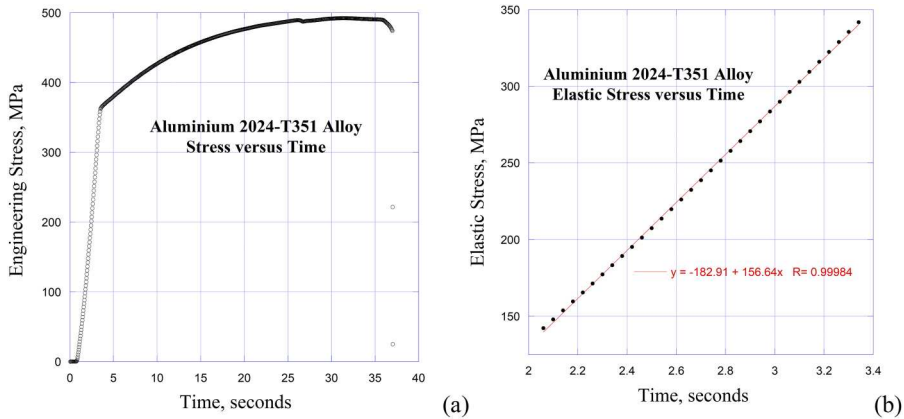

**Figure 2.** Stress vs time for the aluminium alloy. (a) The full curve. (b) The linear elastic loading rate measured as 157 MPa/s. The elastic time is less than 3 s and about a second is needed for the fixtures to fully engage the test bar.

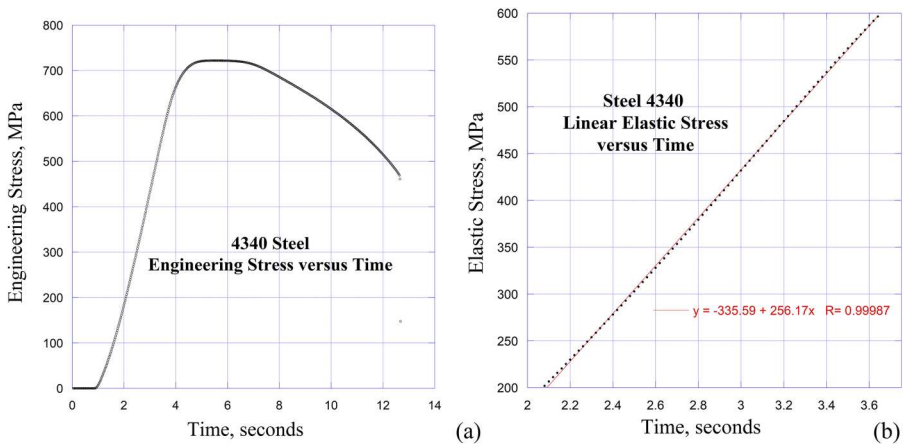

**Figure 3.** Stress vs time for the ferrous alloy. (a) The full curve and (b) the elastic loading rate measured as 256 MPa/s.

Figure 3(a) shows the complete stress versus time curve for a steel test specimen. The steel specimen has an elastic loading rate in Figure 3(b) of 256 MPa/s. The stress versus time is chosen for measurement because the mechanical system's elastic elements are in series and thus the load on each element is identical.

## 2.2. Thermal cooling measurements

The Telops FAST M3k IR camera records temperature at each pixel in the image. The camera's data is accessed using Telops' Reveal IR software on the archived thermal images. The thermal data is processed from each image to obtain thermal maps. The recording array is 256 × 320 pixels, and each pixel has a thermal sensitivity of 20 mK. The Telops IR camera utilises a framing rate of up to 3000 frames per second in full window mode. The camera's image has been software selected to look only at the thermal response from outside of the necked area on the tensile bar. The selected area feature of the Reveal IR software allows the image without the neck to be selected. Figure 4(a) is an image from the IR camera with a thermal picture of the test sample geometry. Figure 4(b) is the record of the temperature versus time recorded for the aluminium sample. Note that the elastic loading part of the stress data is about 2 s in the thermal record while the relief wave from the fracture on the right is significantly faster (as represented by the discontinuity at near 37 s in the curve on the right of Figure 4(b)). The extensive plastic strain after about 22 s, cracks the IR coating so some pieces flake off the sample. The discontinuity at near 37 s is the thermal signature from the fracture relief wave. Figure 4(c) is the thermal record versus time for the steel sample.

Figure 5(a) is the aluminium bar's thermal record from the early elastic part of the curve. The data recorded is quasi-isentropic. The rate of cooling is −0.22 K/s and is taken from the fit to the linear data. Figure 5(b) is the steel sample's elastic thermal cooling record. The steel cooling rate is −0.16 K/s. The times chosen for each sample are the early times that correspond to stresses less than 90% of the yield stresses. Typically, this is less than 2 s for aluminium and steel samples.

## 2.3. Thermal heating measurements

The right side of each thermal graph shows a gap in the temperature as the sample fractures, see Figure 4(b, c) on the right side. The unloading is elastic from the release stress wave generated by the fracture. The specimen's stress is reduced to zero. The areas selected to record these temperature jumps or discontinuities are outside of the necked region. The gage length deformation outside of the neck is uniform. The material outside the neck from Considère's Construction Principle, i.e. $dP = 0$ is elastically unloaded because the load maximum has been passed, the neck is formed, and the load decreases. Thus, the stress in the gage length outside of the neck is decreasing after maximum load.

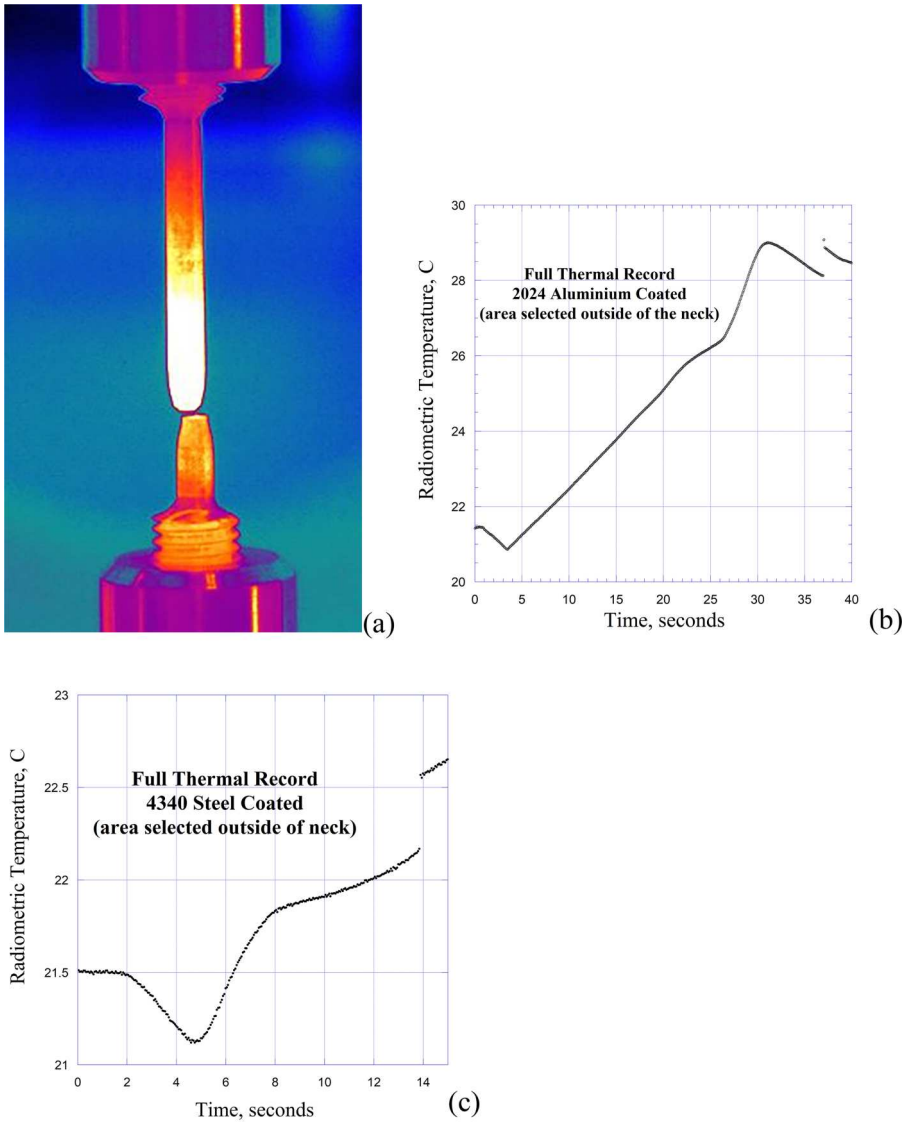

**Figure 4.** (a) Thermal image after fracture in 4340 steel with a slower strain-rate to display plastic heat. The overall thermal image for a steel test bar after fracture; the neck has developed heat as is seen in the image. (b) The full thermal record for a 2024 aluminium bar versus time. The elastic region was completed in about 2.5 s. The bar fractured near 36 s. The irregular behaviour past 22 s to near 35 s may be from the IR coating flaking from the bar. (c) The full thermal record outside of the neck for a 4340-steel specimen versus time. The elastic region was completed in about 2 s. The bar fractured near 14 s.

The thermal heating measurements are based on the true stress, $\sigma$, in the gage length outside of the neck.

$$\sigma = \sigma_e \frac{A_0}{A} \qquad (1)$$

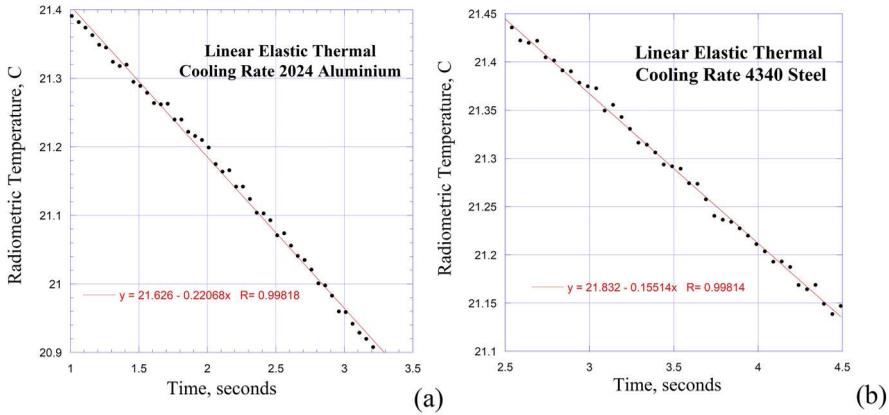

**Figure 5.** Thermal cooling rates in aluminium and steel. (a) The temperature measured versus time in an elastic aluminium bar. The temperature measurement is from the reversible, elastic part of the test. The temperature rate in aluminium is − 0.22°C/s. (b) The temperature rate measured versus time in the elastic part of the steel test bar is −0.16°C/s.

where $\sigma_e$ is the engineering stress, $A_0$ is the initial cross-sectional area of the tensile specimen and $A$ is the true cross sectional area of the tensile specimen outside of the neck at fracture. The engineering stress at fracture when multiplied by the ratio of the areas in Equation (1) gives the elastic stress change, $\Delta\sigma$, upon fracturing. For the test bars used, the ratio of areas in Equation (1) are: steel 101% and aluminium 115%. The temperature of the test specimen is directly recorded at fracture from the thermal record measuring temperature and time. The quasi-isentropic thermo-elastic heating effect is evaluated from $\Delta T$ and $\Delta\sigma$. Fracturing is a fast process, so the release wave response is closer to adiabatic and elastic, but still quasi-isentropic.

### 2.4. Synchronous thermal cooling measurements

The IR camera and the screw driven load-frame are asynchronous instruments. The time records however contain a single reference time that aligns the time axes for both instruments: the instant just before the specimen fractures is common to both data sets. This time reference allows for some data to be a synchronous record of time, stress and temperature. The thermal camera records temperature measurements each 25 ms; the mechanical system records measurements every 20 ms. Thus, for every nine measurements there is one measurement of time, stress and temperature that is synchronous.

The synchronous time as referenced to the fracture of the tensile bar is used to set a starting time for each experiment. The mechanical measurement of zero time is chosen just before the load frame fully engages the tensile bar. The thermal camera's zero time uses the mechanical starting time to record temperature data.

Figure 6 is the synchronous measurements of the temperature versus stress for aluminium. Figure 6(a) shows that the 2024 aluminium tensile bar cools as the stress is increased on the left side of the graph; stress above the yield stress increases the temperature due to plastic deformation. The cooling measured on the left side of the graph is shown in Figure 6(b). The slope is the quasi-isentropic thermoelastic cooling. A linear fit is seen in Figure 6(b) with a slope of −2.1 K/GPa.

Figure 7(a) is the synchronous cooling and heating of the quasi-isentropic data from the 4340 steel test bar. Again, the graph on the left shows temperature cooling due to the application of stress. The increase in temperature curve seen on the right side of Figure 7(a) is the bar heating due to plastic deformation. Figure 7(b) is the linear curve fit of the synchronous cooling temperature versus stress. The slope of −0.75 K/GPa is the quasi-isentropic thermoelastic cooling of the bar. The data in Figures 6(b) and 7(b) are direct physical measurements and do not include any non-linear elements from the machine and fixture compliance.

## 3. Thermodynamic analysis of a tensile test bar

Figure 8 is a schematic of the test bar drawn to define the thermodynamic system used during tensile testing. $P$ is the applied load and $\delta$ is the load-point displacement. The test sample has an initial gage length of $\ell_0$ and a deformed length of $\ell$ as shown in Figure 8. The first and second law's thermodynamic incremental energy balance in the test bar is:

$$dU = TdS + Pd\delta \qquad (2)$$

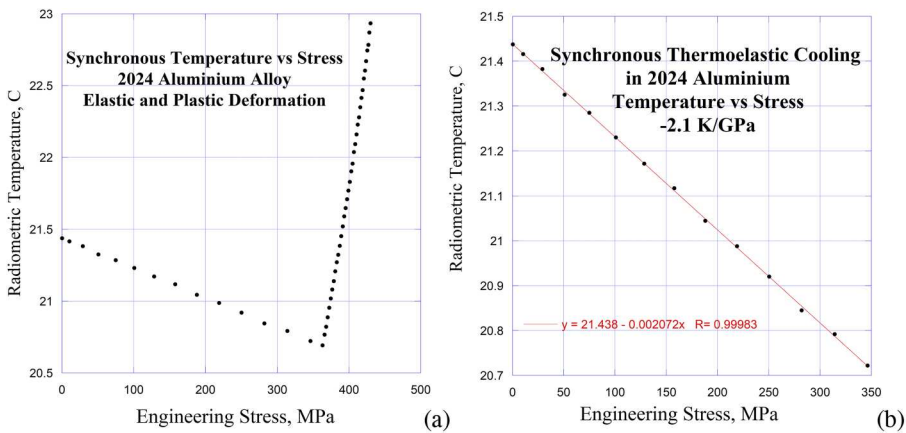

**Figure 6.** The time base was constructed to be synchronous using the instant before fracture as a reference time see text. Synchronous temperature versus stress in 2024 aluminium alloy. (a) The elastic deformation is on the left and plastic deformation is on the right. (b) The thermoelastic curve on the left is fit to a linear relation.

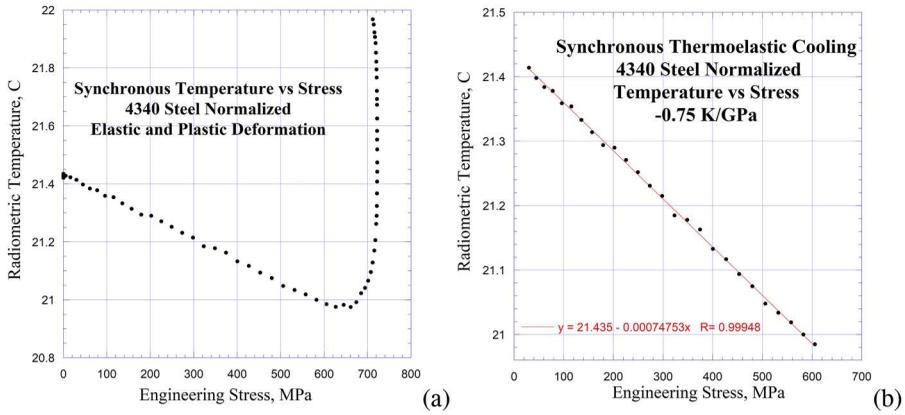

**Figure 7.** Synchronous temperature versus stress in steel. (a) Temperature versus stress in 4340 steel; elastic deformation is on the left and plastic deformation is on the right. (b) The thermoelastic curve on the left is fit to a linear relation.

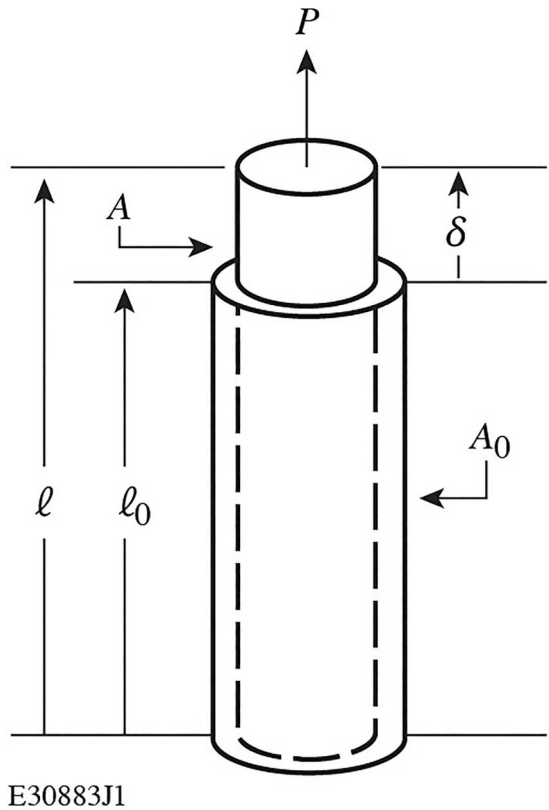

**Figure 8.** A schematic tensile bar shows the geometry, the applied load, $P$ and the load-point displacement, $\delta$, plus the specimen's geometry before and after deformation.

$T$ is the absolute temperature while $S$ is the entropy of the sample. $TdS$ is the incremental heat added to the system and $Pd\delta$ is the incremental work done by the load on the system. The + sign in Equation (2) is because $P$ is a force doing work on the system in the direction of $\delta$. The sum of these two terms is the incremental internal energy change, $dU$. The Gibbs like free energy, $G$ may be found from $U$:

$$G = U - TS - P\delta \tag{3}$$

or

$$dG = -SdT - \delta dP \tag{4}$$

With $T$ and $P$ as the independent variables in the system, the definition of the linear thermal expansion coefficient, $\alpha$ is:

$$\alpha = \frac{1}{\ell}\frac{\partial \ell}{\partial T}\bigg|_P \tag{5}$$

The constant load heat capacity of the sample, $C_P$ is:

$$C_P = T\frac{\partial S}{\partial T}\bigg|_P \tag{6}$$

Appendix A in Equation (A-3) establishes that:

$$\frac{\partial T}{\partial P}\bigg|_S = -\frac{\alpha \ell T}{C_P} \tag{7}$$

$C_P$ is related to the heat capacity per unit mass of the material, $c_p$ by

$$C_P = c_p \rho A_0 \ell_0 \tag{8}$$

with $\rho$ being the material's specific mass density.

$P$ on the left side of (7) is related to the engineering stress, $\sigma_e$ by

$$\frac{P}{A_0} = \sigma_e \tag{9}$$

So, Equation (7) with the aid of (8) and (9) yields a thermodynamic expression connecting changes in $T$ to $\sigma_e$, the elastic stress or see reference [22]:

$$\frac{\partial T}{\partial \sigma_e}\bigg|_S = -\frac{\alpha T}{c_p \rho} \tag{10}$$

The stretch ratio of the bar includes the deformed length, $\ell$ relative to the initial length, $\ell_0$.

$$\frac{\ell}{\ell_0} = 1 + \frac{\delta}{\ell_0} \tag{11}$$

The engineering strain, $e$, is from the deformation, $\delta$, divided by $\ell_0$.

$$e = \frac{\delta}{\ell_0} \tag{12}$$

The engineering strain is small in the elastic range so $\ell$ and $\ell_0$ are nearly equal.

Equation (10) was published in [22] and Thomson's relation [1] was called at that time Lord Kelvin's 'Adiabatic Thermo-Elastic Effect.' Equation (10) is restricted to elastic stress and adiabatic conditions. Experimental measurements from the IR camera and from the MTS instrument have both been asynchronously recorded in our experiments. The temperature and stress changes and their ratio are both recorded versus time, $t$. The sample's fracture time is a reference time so about 10% of our data is presented synchronously. Asynchronous records use $T$ and $\sigma_e$ versus time as seen in Equation (13) on the left to eliminate $t$. The discontinuity in the temperature when the sample fractures is directly measured from the middle term in the expression below. The slope of synchronous data of temperature and elastic stress are also found from the middle expression.

$$\left.\frac{\Delta T}{\Delta t}\right|_S \bigg/ \left.\frac{\Delta \sigma_e}{\Delta t}\right|_S = \left.\frac{\Delta T}{\Delta \sigma_e}\right|_S = -\frac{\alpha T}{c_p \rho} \tag{13}$$

The experimental expression is on the left and in the middle equation while the theoretical value is on the right in Equation (13). This equation permits a direct comparison of the concept of temperature changes due to elastic uniaxial stresses to measured values. The experimental expression in (13) is the quasi-isentropic thermo-elastic stress cooling or heating for uniaxial tensile stresses. For heating the engineering stresses are corrected by Equation (1) for the stress outside the tensile bar's neck. The system measurements have been restricted to the reversible or elastic regions of the tensile loading curves and fracturing curves, so values are quasi-isentropic properties as expanded upon in Section 5.

## 4. Predictions from Sesame and LEOS tables

Equation of state (EOS) tables are most often used in modern radiation-hydrodynamics codes to provide closure to the hierarchy of fluid equations i.e. conservation of mass, momentum, and energy. The purpose of such a table is to be able to determine the material's pressure as a function of mass density and temperature. As such, it is a user-defined modelling choice, and many options exist for the EOS used. Two of the most common choices in the high-energy-density physics community are the SESAME tables [30] and LEOS tables [31, 32] provided by groups at Los Alamos National Laboratory and Lawrence Livermore National Laboratory, respectively. Different EOS model choices include different physical effects and construction choices, such as whether to be thermodynamically consistent and the degree to which discontinuities have been smoothed in the parameter space of density and temperature included in the tables. The main challenge in using EOS tables is numerical accuracy during interpolations to prevent spurious inaccuracies within the simulation. To highlight a case where a ∼30% difference is found between two tables attempting to describe the same material, we include the cooling values obtained from SESAME #2140 and LEOS #260 tables for iron in Table 1.

**Table 1.** Thermoelastic stress cooling and heating experimental and theoretical values.

| Material | Experimental cooling: Measured | Experimental heating: Measured | Synchronous cooling: Measured | Equation (13) Predictions[a] | Sesame tables[b] #3720 Al #2140 Fe | LEOS #260 Iron[b] |
|---|---|---|---|---|---|---|
| Aluminium | −1.7 ± 0.3 K/GPa | −1.8 ± 0.2 K/GPa | −1.8 ± 0.3 K/GPa | −2.8 K/GPa | −2.7 K/GPa | NA |
| Steel | −0.61 ± 0.05 K/GPa | −1.2 ± 0.3 K/GPa | −0.69 ± 0.05 K/GPa | −0.97 K/GPa | −1.2 K/GPa | −0.93 K/GPa |

[a]Table 2 lists material properties of Equation (13) used in Table 1.
[b]Sesame and LEOS tables are for pressures not stresses: $p = -\sigma_e/3$.

### 4.1. Sesame table predictions of thermoelastic stress cooling expressions

Sesame tables were used to evaluate the temperature increase between two points when the pressure difference is known. The points are taken to be on a constant entropy surface. Negative values of pressure in the tables indicate the material is under tension, not compression. In the experiments described above the stresses are in tension as are the values in the tables.

The two states for aluminium are with the first point at $T = 292.4$ K with $p = 0.1$ MPa and the second point is $p = -133$ MPa and $T$ as found from the Sesame Table #3720 as $T = 292.25$ K calculated from the table. The slope is then found on a constant entropy surface and is $\Delta T/\Delta p = -7.95$ K/GPa.

The slope for the steel is again on a constant entropy surface and is evaluated between the states $T = 291.9$ K with $p = 0.1$ MPa and the second point $p = -266$ MPa and $T = 290.93$ K from the Sesame Table #2140. The Sesame table gives a slope of $\Delta T/\Delta p = -3.66$ K/GPa. The values for aluminium and steel are seen in Table 1 on the right side. However, Sesame Tables are for pressures not uniaxial stress. The entries in Table 1 for Sesame for direct comparison to experiments are for uniaxial stress and are only based on the dilatational component of the stress tensor.

### 5. Analysis for thermo-elastic stress cooling and heating

Table 1 is a summary of the experimental and analytic values of thermoelastic stress cooling expressions for aluminium and steel. The experimental data are from the left side of Equation (13). The data for aluminium are from

**Table 2.** Material properties* used to calculate the theoretical values in Table 1.

| Property | Aluminium, 2024 | Steel, 4340 |
| --- | --- | --- |
| Heat capacity, $c_p$ | 0.875 J/g–K | 0.475 J/g–K |
| Specific density, $\rho$ | 2.78 g/cc | 7.85 g/cc |
| Linear thermal expansion coefficient, $\alpha$ | $23.2 \times 10^{-6}$ 1/K | $12.3 \times 10^{-6}$ 1/K |

Figures 2(b), 5(a) and 6(b) plus data from additional samples. For the steel specimens, the data are from Figures 3(b), 5(b) and 7(b) plus data from additional samples. Table 1 also shows the right side of Equation (13) evaluated using the data contained in Table 2. Finally, Table 1 also has expressions evaluated using the published data in Sesame and LEOS tables as already noted.

The values cited above are for uniaxial stresses. The data listed are relevant to temperature versus uniaxial tensile stress. Equation (13) also provides predicted values for comparisons to EOS tables and to cooling and heating measurements.

Data in Table 2 are from ASM Materials Web [29].

The tensile test bars used were optimised for obtaining mechanical properties. The test bars use a significant % of the load cell's capacity so stress may be evaluated to three significant figures. The experiments evaluated are not necessarily ideal for measuring the thermoelastic stress cooling/heating expressions which rely not only on near adiabatic and elastic reversible conditions but also restrictions for heat transfer. Thermal quasi-isentropic measurements in both aluminium and steel would benefit from faster testing and longer gage lengths. The data described here are for quasi-isentropic conditions as described below.

### 5.1. Estimation of errors from assuming quasi-isentropic conditions

Isentropic is an ideal that is difficult to achieve. The term quasi-isentropic is used for all experimental data in the text. There is heat flow in the tensile bars as seen from changes in the thermal images. The heat flow is estimated below by considering an external reservoir that can add or remove heat from the tensile bar. Physically, the reservoir is the threaded ends of the tensile bar, the fixtures holding the bar and the tensile machine. The external reservoir is considered to supply heat or remove heat while remaining isothermal. The bar on initial loading cools and the thermal reservoir supplies heat to the bar. The bar is thus warmer than expected if the test were perfectly isentropic. The plastic deformation as seen in Figures 6(a) and 7(a) produces heat in the bars. After thermoelastic cooling up until the time in the test when the bar's temperature is the same as the reservoir's temperature, heat is added to the test bar. At that point when the bar and reservoir have the same temperature, the heat flow is zero. Plasticity continues to heat the bar until the load reaches an extremum. The deformation before the extremum is homogeneous along the entire gage length of the bar, after maximum load, a neck forms in part of the bar with extensive local plastic deformation. The gage length outside of the neck however is elastic with the stress decreasing. The heat flow into the bar but outside the neck is no longer from plastic deformation as plasticity outside of the neck has ceased; the load and the engineering stress is near an extremum so the thermoelastic heating from stress reduction is near zero and is at zero when $dP = 0$. The heat from the intense deformation in the neck and from the reservoir is flowing into the gage length outside of the neck. The change in temperature per second in the steel bar from both thermal sources are seen in Figure 4(c) just after maximum load. This value is to be compared to the temperature change per second when the thermoelastic effect is being measure at the end and start of the experiments, i.e. both the fracture (heating rate) and during the elastic loading (cooling rate). For the aluminium and steel Table 3 shows these comparisons.

The term quasi-isentropic is used without corrections for temperature rate changes due to heat exchanged with the thermal reservoir. Heating rate corrections are small; cooling rate corrections are more important for the aluminium alloy than for the steel but significant for both materials. The conditions noted in Table 3 are called quasi-isentropic cooling and heating as it establishes temperature changes due to heat flow used in the experiments.

The tensile bar necessarily moves slightly during deformation. One end is attached to the stationary test frame the other end moves with the cross-head of the machine. The IR camera selects a fixed non-moving spatial image to determine the temperature. There is better thermal resolution from the bottom of the tensile bar which is attached to the stationary MTS machine's base. This applies especially to the heating experiments after fracture when the strain in the bar is totally released.

Table 1 is a summary of the thermoelastic stress cooling/heating measurements. The first major comparison is between measured cooling and heating values. The values are not very different although the strain-rates are two orders of magnitude faster in heating. Direct comparisons to Equation (13) are also reasonable. The aluminium measurement is about 65% of the theoretical value. The discrepancy between Equation (13) and these measurements is most likely due to samples straining during testing, emissivity changes and quasi-isentropic temperature changes. The steel more closely agrees with Equation (13) if the cooling and heating values are added for comparison to the theoretical value. This comparison is about 95%. Yet, the complex thermal behaviour observed in the complete tensile tests show the specimens cool elastically, then heat up due to plasticity, heat up further after the load reaches a maximum by elastic unloading outside of the neck and again heat up further upon fracture by stress release. Finally, it slowly heats with energy from the intense plastic neck leaking heat into the material outside the neck and slowly cools with thermal transfer into the load frame. This behaviour cautions about heat transfer and fully adiabatic conditions during cooling as shown in Table 3. The heating measurements due to stress release by fracture are quite fast. The measurements recorded, the uniaxial energy balance values and Sesame plus LEOS values are given in Table 1 for comparisons. There are also reasonable agreements with the Sesame and LEOS tables.

Table 3. Heating and cooling rate comparisons to temperature rate change with $dP = 0$.

|  | Heating rate | Cooling rate | Heating rate at $dP = 0$ |
| --- | --- | --- | --- |
| Aluminium | 25°C/s | −0.22°C/s | 0.16°C/s |
| Steel | 22°C/s | −0.16°C/s | 0.04°C/s |

### 5.2. Role of shear stress in cooling and heating analysis

In solids, two phonons in three are shear phonons, the other is a longitudinal phonon. Yet, the shear energy is only apparent through the heat capacity in Equation (13). A uniaxial stress state contains significant shear stress. However, no additional temperature changes from the deviatoric components of the stress tensor are apparent in the results presented above. Thermal shear stress effects are explored below using octahedral stresses. Octahedral stresses favour expressions of pressure and shear for representing the general stress tensor.

Equation (13) only includes elastic temperature changes from shear strain energy in the specific heat capacity as noted above. If shear were to be directly included, then the thermodynamic system would need three independent state variables. The addition of shear contributions to tensile test mechanics is very important as shear is the main basis for the nonlinear responses seen in Figures 2(a) and 3(a). Shear is also the basis for heating after yield but before fracture. From a thermodynamic perspective shear properties are rarely found in the literature for the thermoelastic mechanical response. So, why is shear not well represented in the material's thermal response? As seen in the more complex octahedral stress state [27, 33, 34] noted below, it is obvious that shear makes no contribution to the temperature changes in Equation (13).

The state of stress in a uniaxial test is shown on the left in Equation (14). The middle and right expressions are from a tensor rotation: the rotation is 54.71° of the principal stress axis 11 about a line that is at 45° to the 22 and 33 axes; it bisects the 22 and 33 axes. This is the first octahedral plane.

$$\begin{pmatrix} \sigma_{11} & 0 & 0 \\ 0 & 0 & 0 \\ 0 & 0 & 0 \end{pmatrix} \Leftrightarrow \begin{pmatrix} \frac{\sigma_{11}}{3} & \frac{\sigma_{11}}{3} & \frac{-\sigma_{11}}{3} \\ \frac{\sigma_{11}}{3} & \frac{\sigma_{11}}{3} & 0 \\ \frac{-\sigma_{11}}{3} & 0 & \frac{\sigma_{11}}{3} \end{pmatrix} \Leftrightarrow \begin{pmatrix} -p & \tau & -\tau \\ \tau & -p & 0 \\ -\tau & 0 & -p \end{pmatrix} \quad (14)$$

The uniaxial stress tensor describes an octahedral stress state in the middle and right expressions. The pressure, $p$, is – 1/3 of the trace of the stress diagonal and the shear is the octahedral shear, $\tau$.

$$p = -\frac{\sigma_{11}}{3} \text{ and } \tau = \frac{\sigma_{11}}{3} \quad (15)$$

Equation (14) shows that when Equation (10) or (13) describes the thermal changes due to stresses then the two shear terms containing $\tau$ on the right in Equation (14) will make no contribution to temperature changes in the elastic test bar. This conclusion is based on all shear thermal expansion coefficients being zero. Recent theoretical predictions [35] contradict the concept of sheared solids having no thermal expansion coefficients. It was also recently demonstrated [27] that in material which includes both pressure and shear that when the pressure is held constant, shear stresses contribute to the isothermal ratio of heat to work. That ratio, $R$, is found from,

$$R = \frac{T\partial S}{2v\tau\partial\gamma}\bigg|_{T,p} = \frac{\partial \ell n(\text{shear compliance})}{\partial \ell n(T)}\bigg|_{p,\tau} \quad (16)$$

In expression (16) $\gamma$ is the shear strain, $v$ is the specific volume and $R$ is a material specific number empirically between 0 and about 2 that depends only on the temperature. The temperature dependence of the shear compliance, $\lambda$, is seen on the right in Equation (16). $\lambda$ is the reciprocal of the shear modulus. Expression (16) evaluated for aluminium has $R_{Al} = 0.187$ from the data of [36] and $R_{Fe}$ for steel is 0.059 taken from the data of [37, 38]. The isothermal heat found in $R$ is blocked in material that is sheared isentropically so the heat will not enter (or leave) the system. Thus, the material will necessarily cool (or heat) by exchanging energy through the heat capacity. The following expression determines the isentropic temperature change, $\Delta T$, of the material using the heat capacity:

$$c_p \Delta T = -R \int 2v\tau d\gamma \quad (17)$$

The change in temperature in expression (17) is added to the dilatational stress driven temperature change from Equation (13). A uniaxial stress state has shear stresses in both tension and compression. Thus, shear cools material in uniaxial compression while in uniaxial compression, the stress component will heat the material as seen from Equation (13). Thus, from Equation (13) plus (17) we have the total temperature change in an isentropic system, i.e. elastic material as:

$$\Delta T = \left(-\frac{\alpha T}{c_p \rho}\Delta \sigma - \frac{R \int 2v\tau d\gamma}{c_p}\right) = -\frac{\alpha T}{c_p \rho}\Delta \sigma - R\frac{\lambda \tau^2}{c_p \rho} \quad (18)$$

The second term on the right is from the shear stress in the material and is always negative. $\tau^2$ in Equation (18) guarantees that the second term is negative. For our aluminium specimens, we estimate the values in Equation (18) assuming we have uniaxial compression, to be for materials that are at the yield stress:

$$\Delta T = 1.10\text{K} - 0.02\text{K} = 1.08\text{K} \quad (19)$$

The first term is from compressive stress and the second is from shear. The shear term at this stress level is thus small.

For our steel specimens, we estimate the values in Equation (18) to be:

$$\Delta T = 0.77\text{K} - 0.053\text{K} = 0.72\text{K} \quad (20)$$

The second terms in Equations (19) and (20) are due to shear. In the case of aluminium, with $R_{\text{Al}} = 0.187$ the second term is quite small compared to the first term. In the case of the steel, with $R_{\text{Fe}} = 0.059$ the shear term is also significantly smaller than the first term. The effect of shear in high-strength, light alloys could be more significant. The effects of shear are below our measurement capabilities in both materials. At very high pressures of say 1 TPa as in laser ablated, compressed materials [39–41], the opposite conclusion would be reached.

The metallic specimens tested are well described by a tensile stress without shear to find the temperature changes. Some ceramic materials with limited plasticity might be expected to show more effects from shear. For example, diamonds in laser driven longitudinal uniaxial strain experiments might have sufficiently high yield stresses, without shear yielding being dominant. If the yield stress, $\sigma_y$, were:

$$\sigma_y > \frac{\alpha T}{R\lambda} \quad (21)$$

Then shear effects in the thermodynamic system would be important. Shear [35] gives a very different thermodynamic description of a material's behaviour when compared to pressure effects alone. The right side of expression (21) is 2.08 and 9.75 GPa for 2024 aluminium and 4340 steel respectively. These yield stress values are well above the values measured in Figures 2(a) and 3(a).

## 6. Conclusions

Direct temperature measurements of uniaxial stressed bars have been recorded and analysed. The effect is quasi-isentropic thermoelastic stress cooling or heating depending on the stress sign. Direct comparisons between experimental and derived properties [9] in solids are rare. The comparisons between experimental measurements and theoretical predictions given in Table 1 establish basic agreement among heating, cooling, synchronous cooling, equation of state tables, Thomson's thermoelastic prediction and the temperature-engineering stress based thermo-elastic analysis. The experimental values are remarkably well described by an analytic thermodynamic expression from the 1850s. The experimental data are closely described by equations of state tables even though the pressures and temperatures are very modest for both Sesame and LEOS tables. The Sesame and LEOS tables support Thomson's predictions and the quasi-isentropic thermoelastic temperature changes measured.

The isentropic temperature cooling and heating equality demonstrates that the isentropic temperature changes are recovered when the stress is relieved. The results observed were from temperature cooling in tensile elastic material which was then fully recovered by heating when the stress was released. Thermal recovery from the release stress wave was unexpected and is the first study of temperature heating by stress relief. It was observed after extensive plastic deformation had significantly heated the material, the fracture stress when released added a temperature increase that recovered the temperature from cooling imposed by the elastic stresses. It is only the quasi-isentropic temperature that is recovered. The temperature increase from plasticity is irreversible. The temperature change measurements are small, but they are universal. The temperature restoration by the quasi-isentropic stress release wave was always observed after the tensile bar fractured. The specimen's strain-rate from the fracture relief wave is two orders of magnitude higher than the strain-rate in the tensile test and is independent of the loading rate. Temperature recovery from the release waves is with a quasi-isentropic restriction on the material and these temperature observations and measurements are new. They also will apply to all rapid, elastic stress relief in solids.

Our experimental evidence supports a simple two variable thermodynamic systems using $T$ and $P$ as seen in Equation (13) and Table 1. The two variable model proposed supports thermodynamic changes to the temperature from the adiabatic constraint and dilatational expansion. Thomson's model uses temperature and all six stresses. Shear stresses which can only be described with additional mechanical state variables, have no observable thermoelastic effects; thermal heating from plastic deformation is seen in all the full temperature plastic measurements.

The predicted thermal shear terms in our elastic measurements are small, well below what we can easily experimentally measure. The elastic effects of normal and shear stresses are very different: the thermal behaviour from normal stresses is proportional to the stress while thermal changes from shear are proportional to the stress squared. The temperature responses due to elastic stresses in our study are totally dominated by normal stresses at the stress levels in the experiments.

### Acknowledgements

SJB would like to thank Andrea Pickel for early discussions on the difficulty of thermal measurements in tensile tests. He would also like to thank Wes Autran who helped with providing the Telops IR camera used in the measurements. Renato Perucchio has graciously made available the facilities to complete this research. A reviewer has provided many important suggestions and urged the authors to obtain the synchronous data files; we are very thankful for the reviewer's persistence. For ABS this material is partially based on work supported by the United States Department of Energy under award DENA0003856.

## Disclosure statement

No potential conflict of interest was reported by the author(s).

## ORCID

*S. J. Burns*  http://orcid.org/0000-0003-4905-3130

## Appendix A

Derivation of Equation (10) using Jacobian algebra with the definitions of linear thermal expansion, $\alpha$, from Equation (5), heat capacity at constant load, $C_P$, from Equation (6) and the specimen compliance at constant $T$, $\lambda_T$.

A Jacobian Table with temperature, $T$ and load, $P$ as the independent variables for the system in Figure 8.

**Table A1.** Jacobian table for a uniaxial tensile bar.

| Thermodynamic state variable | $\left.\frac{\partial}{\partial T}\right|_P$ | $\left.\frac{\partial}{\partial P}\right|_T$ |
|---|---|---|
| Temperature, $T$ | 1 | 0 |
| Load, $P$ | 0 | 1 |
| Entropy, $S$ | $\frac{C_P}{T}$ | $\alpha \ell$ |
| Load point displacement, $\delta$ | $\alpha \ell$ | $\lambda_T$ |
| Gibbs function, $G$ | $-S$ | $-\delta$ |

Use has been made of Equation (4) for the incremental Gibbs function. The isothermal specimen compliance is defined as

$$\lambda_T \equiv \left.\frac{\partial \delta}{\partial P}\right|_T \quad (A-1)$$

And the Maxwell relation seen in A-2, follows from Equation (4). It is used to connect $S$ to $\delta$ in Table A1.

$$\left.\frac{\partial S}{\partial P}\right|_T = \left.\frac{\partial \delta}{\partial T}\right|_P \quad (A-2)$$

It follows that:

$$\left.\frac{\partial T}{\partial P}\right|_S = \frac{J(T,S)}{J(P,S)} = -\frac{\alpha \ell T}{C_P} \quad (A-3)$$

Equation (A-3) with the aid of Equations (8) and (9) gives Equation (10).